\documentclass{article}

\usepackage{multirow}
\usepackage{arxiv}
\usepackage{graphicx}
\usepackage[utf8]{inputenc} 
\usepackage[T1]{fontenc}    
\usepackage{hyperref}       
\usepackage{url}            
\usepackage{booktabs}       
\usepackage{amsfonts}       
\usepackage{nicefrac}       
\usepackage{microtype}      
\usepackage{lipsum}		
\graphicspath{{Figure/}}

\def\eg{{\em e.g.}}
\def\ie{{\em i.e.}}

\title{Microscope Based HER2 Scoring System}

\author{
  Jun~Zhang \\
  Tencent AI Lab\\
  Shenzhen, Guangdong,\\
  China \\
  \texttt{junejzhang@tencent.com}
   \And
  Kuan~Tian\\
  Tencent AI Lab\\
  Shenzhen, Guangdong,\\
  China \\
  \texttt{kuantian@tencent.com}
  \And
  Pei~Dong\\
  Tencent AI Lab\\
  Shenzhen, Guangdong,\\
  China \\
  \texttt{peidong@tencent.com}\\
  \And
  \And
  Haocheng~Shen\\
  Tencent AI Lab\\
  Shenzhen, Guangdong,\\
  China \\
  \texttt{hcshen@tencent.com}
  \And
   Kezhou~Yan\\
  Tencent AI Lab\\
  Shenzhen, Guangdong,\\
  China \\
  \texttt{kezhouyan@tencent.com}
  \And
  Jianhua~Yao\\
  Tencent AI Lab\\
  Shenzhen, Guangdong,\\
  China \\
  \texttt{jianhuayao@tencent.com}
  \And
  Junzhou~Huang\\
  Tencent AI Lab\\
  Shenzhen, Guangdong,\\
  China \\
  \texttt{joehhuang@tencent.com}
  \And
  Xiao~Han\\
  Tencent AI Lab\\
  Shenzhen, Guangdong,\\
  China \\
  \texttt{haroldhan@tencent.com}
}

\begin{document}

\maketitle
%
%
\begin{abstract}
The overexpression of human epidermal growth factor receptor 2 (HER2) has been established as a therapeutic target in multiple types of cancers, such as breast and gastric cancers. Immunohistochemistry (IHC) is employed as a basic HER2 test to identify the HER2-positive, borderline, and HER2-negative patients. However, the reliability and accuracy of HER2 scoring are affected by many factors, such as pathologists' experience. Recently, artificial intelligence (AI) has been used in various disease diagnosis to improve diagnostic accuracy and reliability, but the interpretation of diagnosis results is still an open problem. In this paper, we propose a real-time HER2 scoring system, which follows the HER2 scoring guidelines to complete the diagnosis, and thus each step is explainable. Unlike the previous scoring systems based on whole-slide imaging, our HER2 scoring system is integrated into an augmented reality (AR) microscope that can feedback AI results to the pathologists while reading the slide. The pathologists can help select informative fields of view (FOVs), avoiding the confounding regions, such as DCIS. Importantly, we illustrate the intermediate results with membrane staining condition and cell classification results, making it possible to evaluate the reliability of the diagnostic results. Also, we support the interactive modification of selecting regions-of-interest, making our system more flexible in clinical practice. The collaboration of AI and pathologists can significantly improve the robustness of our system. We evaluate our system with 285 breast IHC HER2 slides, and the classification accuracy of 95\% shows the effectiveness of our HER2 scoring system.

\end{abstract}

\section{Introduction}

Human epidermal growth factor receptor 2 (HER2) overexpression has been discovered implicative in the development of multiple types of cancers (\eg, breast and gastric cancers). For example, HER2 is a protein that promotes the growth of breast cancer cells, and it is overexpressed in around 20-30\% of breast cancer tumors~\cite{mitri2012her2}. The amplification or overexpression of this oncogene can cause aggressive breast cancer, which has a high recurrence rate and short survival. Therefore, the HER2 protein has been employed as a significant biomarker/target of selecting a proper therapy plan (\eg, trastuzumab therapy) for breast cancer patients.


Immunohistochemistry (IHC) HER2 test is a basic test to see whether a cancer cell has too much of the HER2 receptor protein on the surface (i.e., membrane). Currently, the HER2 score is non-quantitatively (or at most half-quantitatively) provided by pathologists according to membranous staining of the HER2 protein. The score is given from 0 to 3+ based on the scoring guidelines, such as the American Society of Clinical Oncology and the College of American Pathologists (ASCO/CAP) guidelines for breast cancer (as shown in Table~\ref{table_guideline}). However, visual estimation usually has the problem of intraobserver/interobserver variability. 


Artificial intelligence (AI) is widely used in disease analysis. Many studies have approved that end-to-end training for disease diagnosis is efficient and accurate~\cite{zhang2019pathologist,campanella2019clinical,qaiser2019learning,lian2020hierarchical}. However, the interpretation of diagnostic results is still an open problem. For some special applications that the diagnoses of the disease have clear guidelines, it would be more convincing to calculate the clinically defined measurements for the quantitative analysis. HER2 scoring is one specific clinical application that has clear guidelines to follow to complete the diagnosis.

Currently, there are two ways of using AI to assist pathologists. One solution is utilizing the whole-slide images that are typically acquired at magnifications of $20\times$ or $40\times$, generating gigapixel two-dimensional images that are challenging for computer calculation in an exhaustive manner. Also, it is not easy to distinguish some confounding regions, such as ductal carcinoma in situ (DCIS), from infiltrating carcinoma without seeing H\&E and other immunohistochemistries (\eg, p63, Calponin, SMMHC, SMA, and CD10). Differently, with the use of real-time imaging of microscope, pathologists can help AI system to select typical fields of view (FOVs). In return, pathologists can further benefit from the AI-assisted microscope by presenting both qualitative cell/membrane illustration and quantitative statistics. Furthermore, the interaction between pathologists and AI helps achieve accurate and robust diagnostic results. Notably, fully automatic may not that necessary in current clinical practice.

In this paper, we propose a real-time HER2 scoring system based on an augmented reality (AR) microscope for breast cancer, which follows the breast HER2 scoring guidelines (see Table~\ref{table_guideline}) to make the diagnosis, and thus each step is explainable. Besides analyzing the staining membrane only, we perform a cell-level classification to accomplish the scoring target. Importantly, we illustrate the intermediate results with membrane staining condition and cell classification results, making it possible to evaluate the reliability of the diagnostic results. Our system also supports the interactive modification of tumor regions and a slight adjustment of cell classification results, improving its robustness and flexibility in the clinical application.
Overall, our system has the following features.
\begin{itemize}
  \item Our scoring system is integrated into the microscope, which adapts the existing workflow of the pathologists.
  \item Our diagnostic results are obtained according to the current HER2 scoring guidelines, which has legible interpretability.
  \item Besides membrane delineating, our AI results also contain cell-level detection and classification, which provides accurate quantitative analysis.
  \item The intermediate results of membrane extraction and cell classification are visually illustrated to the pathologists, which can be used to evaluate/rectify the correctness of results.
  \item Our system also supports interactive modification, which makes our system more robust in clinical practice.
\end{itemize}
\begin{table}[!t]

\caption{Breast HER2 scoring guidelines.}
\label{table_guideline}
\centering
\begin{tabular}{l|c|c}
\hline
\hline
HER2 score&Staining condition& Proportion \\
\hline
IHC 3+ (positive) & Circumferential membrane staining that is complete and intense& $\geq 10\%$ of tumor cells\\
IHC 2+ (equivocal) & Weak to moderate complete membrane staining & $\geq 10\%$ of tumor cells\\
IHC 1+ (negative) & Incomplete membrane staining that is faint/barely  & $\geq 10\%$ of tumor cells\\
IHC 0 (negative) & No staining/membrane staining that is incomplete and is faint/barely  &$< 10\%$ of tumor cells\\
\hline
\hline

\end{tabular}
\end{table}

\begin{figure}[t]
	\centering
	\includegraphics[width=\textwidth]{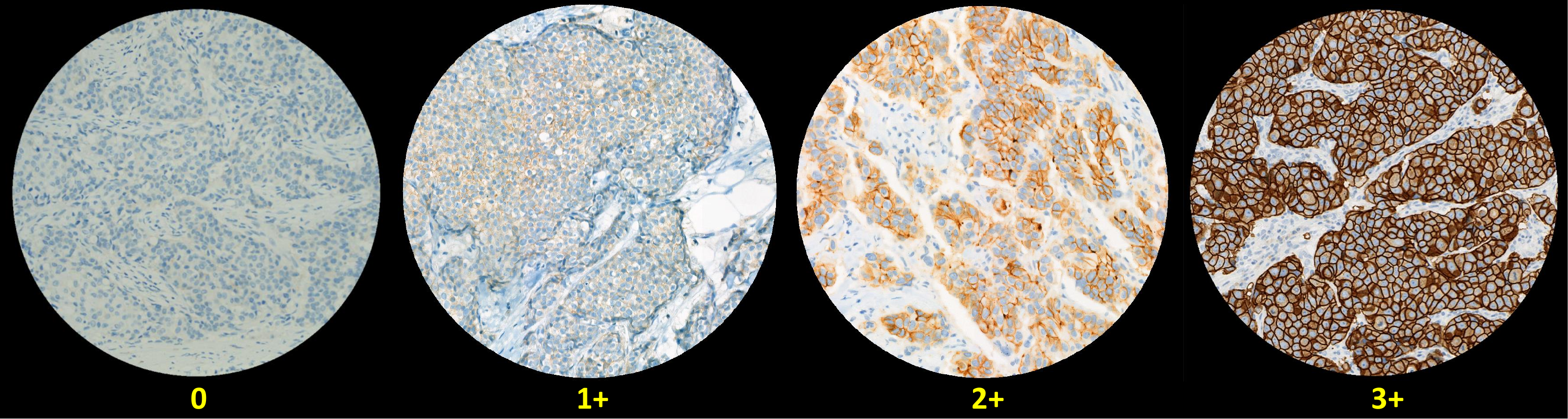}
	\caption{Typical views in breast cancer HER2 slides with different HER2 scores. }
	\label{fig_example}
\end{figure}

\section{Related Work}
As shown in Table~\ref{table_guideline}, ASCO/CAP breast HER2 interpretation guidelines classify the slides into four scores (\ie, 0, 1+, 2+, and 3+), according to the staining patterns of the membrane and the proportions of these staining patterns. The scores of 1+/0, 2+, and 3+ are defined as HER2-negative, borderline, and HER2-positive, respectively. The existing HER2 scoring methods achieve the classification task using two types of categories, including conventional image processing-based and machine-learning-based methods (especially for deep learning). 

\subsection{Methods based on conventional image processing}
A typical method for HER2 scoring is the ImmunoMembrane~\cite{tuominen2012immunomembrane}. The method performs the HER2 scoring by identifying the completeness and area proportion of the staining membrane. The membrane staining condition is transformed into an IM-score for further classification (0/1+, 2+, or 3+). Another software named HER2-CONNECT$^{\rm TM}$~\cite{brugmann2012digital} was developed for HER2 scoring based on the connectivity of membrane in region-of-interest (ROI). Very high overall percentage agreement between automated scoring and ground-truth labels was achieved. There are also some other studies using computer-aided diagnosis software for calculating quantitative measures~\cite{matkowskyj2000quantitative,lehr2001quantitative,hatanaka2001quantitative}. Most of the methods rely on manual intervention (\eg, delineating regions-of-interest). Besides, some commercial analysis systems (\eg, BLISS from Bacus Laboratories, ACIS from Clarient, and ScanScope from DakoCytomation) are developed for the scoring of HER2. With such quantitative software, studies have performed quantitative digital image analysis for HER2 ROIs/WSI, and the results suggested that such quantitative measures have good concordance with experienced pathologists' read and could reduce HER2 IHC equivocal (2+) cases~\cite{li2020quantitative,brugmann2012digital, holten2015optimizing,laurinaviciene2011membrane, helin2016free}.  However, these methods analyze the stained membranes only, and the scoring strategies is different from the criteria in HER2 scoring guidelines.

\subsection{Methods based on machine learning}
Masmoudi et al. proposed to extract quantitative features (\ie, membrane completeness and average membrane intensity) related to HER2 membrane staining for ROI images, and a minimum cluster distance (MCD) classifier is employed to construct the automated IHC assessment of HER2 score in the breast cancer tissue WSI~\cite{masmoudi2009automated}. Vandenberghe et al. developed an automated HER2 scoring method based on deep learning~\cite{vandenberghe2017relevance}. The cancer cell types (\ie, immune cells, stromal cells, artifacts, tumor 0 cells, tumor 1+ cells, tumor 2+ cells, and tumor 3+ cells) are automatically classified by neural network model. Thus, the HER2 scores could be estimated with the percentage of these cells. However, accurate annotations of these different types of cells are challenging. Saha et al. proposed a Her2Net for both nuclei and membrane segmentation, followed by a HER2 score classification~\cite{saha2018her2net}. However, the method was evaluated in small patch-level image classification. Khameneh et al. developed a deep learning framework for WSI image classification~\cite{khameneh2019automated}. First, the epithelial tissue is identified using a superpixel-based support vector machine (SVM) classifier. Then, a neural network (\ie, U-Net) is employed for membrane segmentation. Finally, the overall score based on intensity and completeness of the WSI can be calculated to perform the final classification of HER2 scores. Qaiser et al. presented a method based on deep reinforcement learning to simulated a sequential learning task~\cite{qaiser2019learning}. The model can identify diagnostically relevant ROIs and then learn discriminative features, and the next relevant location is estimated as well. These methods achieve state-of-the-art scoring performance on WSIs (or ROIs from WSIs). However, very few HER2 scoring systems were designed for an AI-assisted microscope so far.


\section{Microscope HER2 Scoring System}

\begin{figure}[t]
	\centering
	\includegraphics[width=\textwidth]{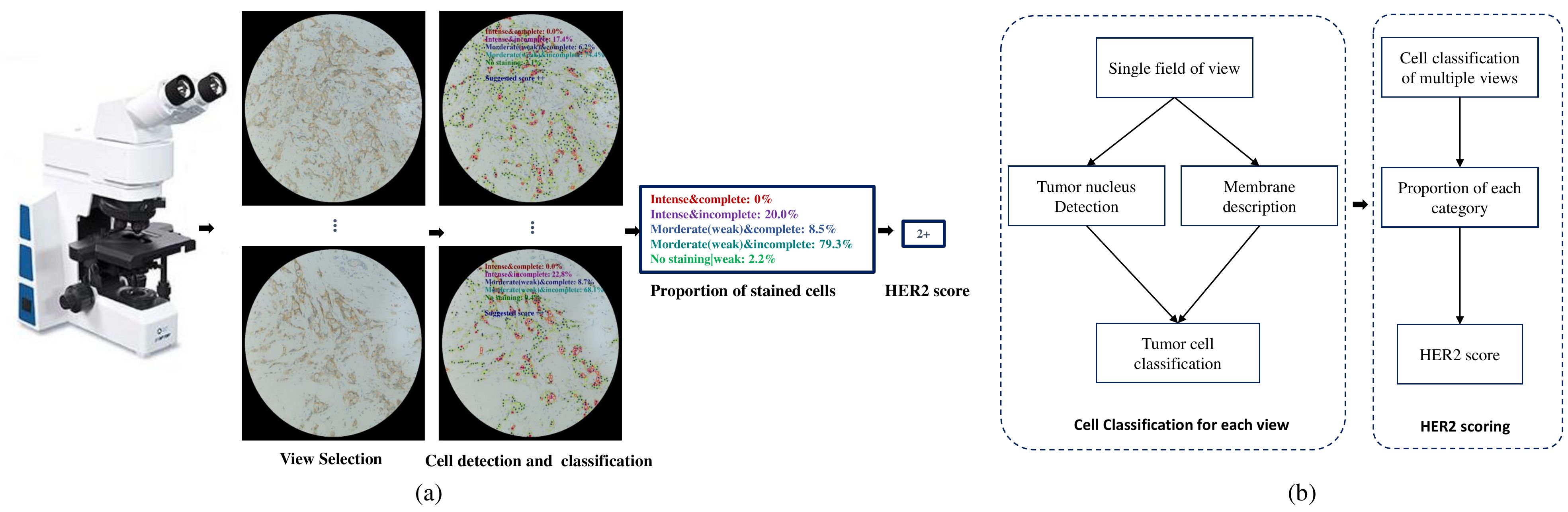}
	\caption{Framework of our microscope-based HER2 scoring system. }
	\label{fig_sys}
\end{figure}

Our algorithm is integrated into an augmented reality (AR) microscope. Similar to the AR microscope developed by Google~\cite{chen2019augmented},  our microscope has the AR display and screen display. As shown in Fig.~\ref{fig_sys} (a), the pathologist reads the slide normally, stays in the field of view that needs to see the AI results, and presses the foot pedal button (returns a signal to the computer to start the calculation). Then, the computer feeds back the calculated results to the AR screen and the computer screen in real-time. The results of multiple typical FOVs are aggregated to calculate the HER2 score.

To obtain a convincing interpretation of the image, we perform the HER2 scoring strictly following the diagnostic guidelines. As shown in Fig.~\ref{fig_example}, IHC HER2 slides can monitor the cell membrane overexpression. The IHC provides a score from 0 to 3+ that measures the amount of HER2 proteins on the membrane of cells in a breast cancer tissue sample.  Generally, the tumor cells with HER2 overexpression are stained with diaminobenzidine (DAB) that is shown in brown, and all nuclei are stained in blue. In order to complete the HER2 scoring according to the diagnostic guidelines, we should identify the tumor cells and perform HER2 scoring according to the membrane staining condition (\ie, intensity and completeness).

Specifically, as shown in Fig.~\ref{fig_sys}, the AI results for each view includes the detection and classification results in tumor cells.  We employ a deep learning algorithm for tumor nucleus detection and utilize the threshold-based segmentation method to segment the cell membranes. The cell membranes are further rectified into several staining masks by image skeletonization and morphological operation. Finally, the cells are classified into five categories according to the associations between the detected nuclei and the staining masks. The AI results contain the recognized cells (shown in distinct color points for each category) and contours of membranes (shown in distinct color lines for the complete and incomplete cell membranes).  After receiving AI results from multiple views, the proportion of each cell category is employed to calculate the HER2 score according to the HER2 guidelines.


\subsection{Training data preparation}
In order to train a robust and accurate tumor nucleus detection model, cell-level point/bounding box annotations are necessary. However, manually annotating nuclei in images are labor-expensive. For example, there are usually thousands of nuclei in an image captured in the view of 20 power objective. Therefore, we propose to use a semiautomatic annotation method from scratch. We first employ an image-processing-based method to extract rough tumor nuclei from images with the simple filtering strategies.

\begin{figure}[t]
	\centering
	\includegraphics[width=\textwidth]{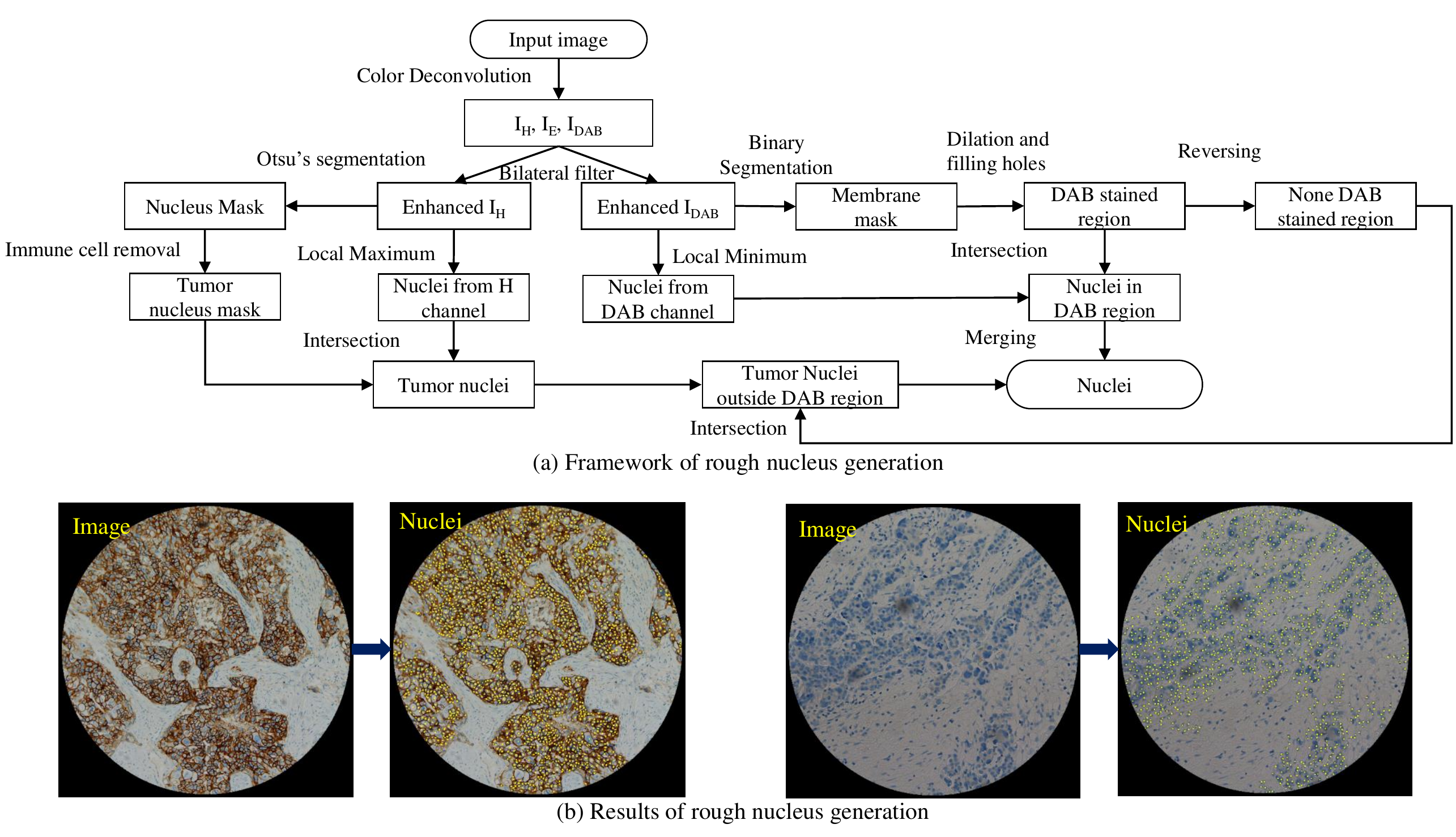}
	\caption{Initial rough nucleus generation based on image processing. }
	\label{fig_nuclei}
\end{figure}

For the standard staining, even if cancer cells are stained with intense/moderate membranes, stromal/immune cells will not have brown membrane staining. As shown in~\ref{fig_nuclei} (a), we would like to detect the nuclei for regions with and without stained membranes separately. With a color deconvolution method~\cite{ruifrok2001quantification}, we can rectify the RGB image to Haematoxylin-Eosin-DAB (HED) color space.
Then, the H channel and DAB channel images (\ie, $\rm\textbf{I}_H$, and $\rm\textbf{I}_{DAB}$) are filtered and enhanced separately by bilateral filtering. Bilateral filtering is a nonlinear filter that can achieve the effects of maintaining edges and reducing noise and smoothing. With the enhanced images, we can identify the nuclei by finding local maximums in $\rm\textbf{I}_H$ and finding local minimums in $\rm\textbf{I}_{DAB}$. For the intensive DAB stained region, we would like to use the nuclei from the DAB channel and nuclei from H channel otherwise. However, nuclei from H channel are the nuclei from not only tumor cells but also from stromal/immune cells. Therefore, we perform simple segmentation and remove stromal/immune nuclei by removing nuclei with small areas. Finally, we can merge the two types of nuclei (from $\rm\textbf{I}_H$ and $\rm\textbf{I}_{DAB}$) by whether they belong to the DAB regions. Fig.~\ref{fig_nuclei} (b) shows the nucleus detection results for two images with intense staining and no staining, respectively. As shown in the figure, most of the nuclei can be correctly detected.

With the initial coarse locations of tumor nuclei, pathologists also manually correct these detected results with our online annotation tool, as shown in Fig.~\ref{fig_annotation} (a), where false positive detections will be removed, and missing nuclei will be added. By doing so, we will receive plenty of effective cell-level point annotations in a short period.



\subsubsection{Heatmap Regression}
\begin{figure}[t]
	\centering
	\includegraphics[width=\textwidth]{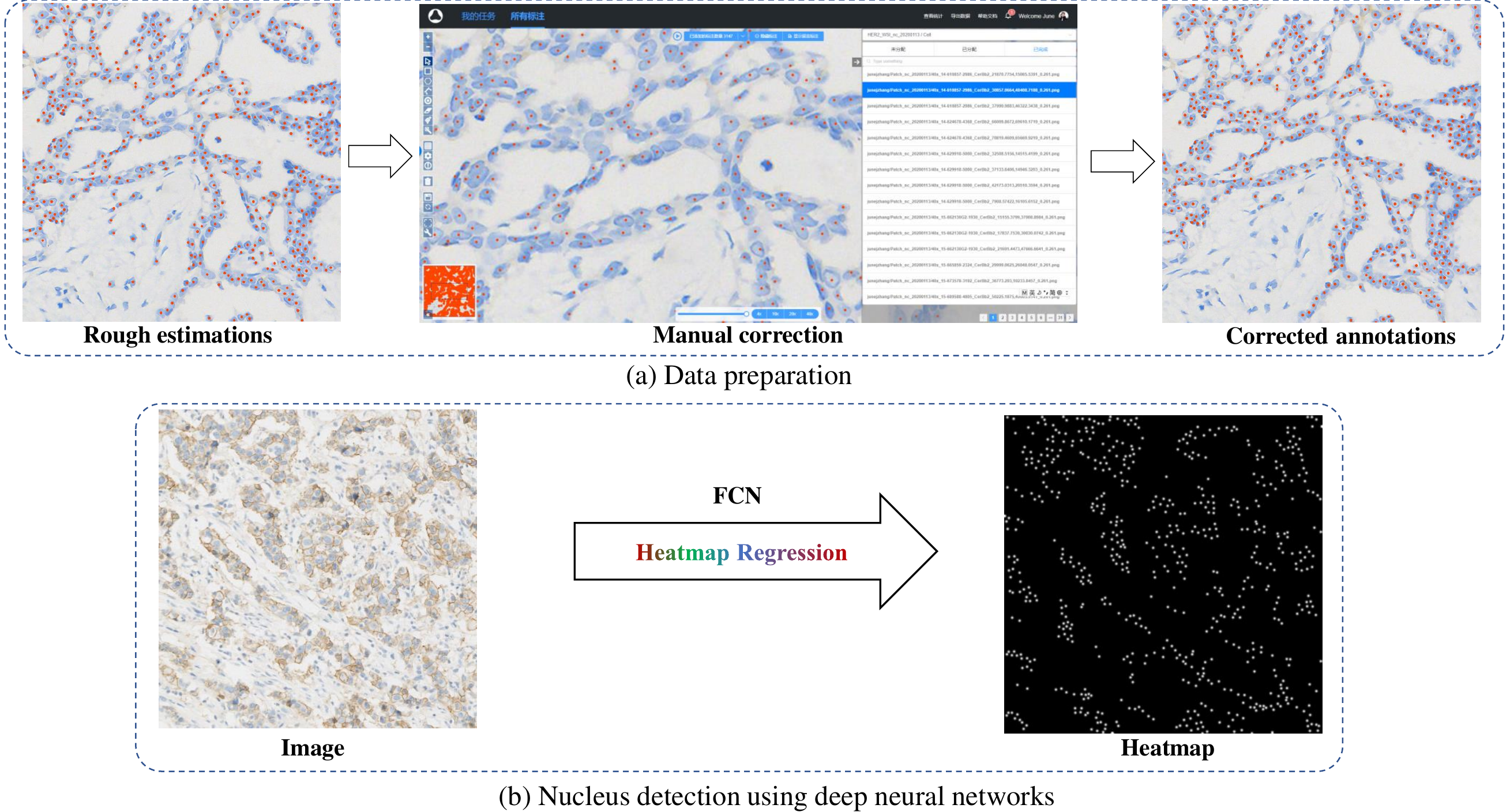}
	\caption{Nucleus detection using deep neural networks. }
	\label{fig_annotation}
\end{figure}
Currently, there are many object/instance detection algorithms can be used for nucleus detection. In our algorithm, we employ the heatmap regression using fully convolutional networks (FCN) in an end-to-end manner~\cite{payer2016regressing,zhang2020context}. Specifically, the RGB image is input, and the output is a heatmap, where each nucleus has a Gaussian-like response centered on the nucleus (as shown in Fig.~\ref{fig_annotation} (b)). The mean square error constraint is used as the loss function to train the model. In the inference stage, all tumor nuclei can be identified by finding the local maximum response position of the predicted heatmap. To facilitate subsequent cell classification, we denote the set of detected nuclei as $\mathbb{D}_{\rm detected}$.

\subsection{Tumor Membrane Description}
Our membrane extraction is similar to the existing methods based on the segmentation of the stained membrane and skeletonization of the segmentation mask. With the extracted contours by skeletonization, the completeness of the membrane can be identified. Different from previous methods, we extract several segmentation masks to identify the cells with varying styles of stains(intense/weak in membrane intensity and complete/incomplete in membrane contours).
\subsubsection{Membrane Segmentation}
We segment the membrane in the DAB channel image using a thresholding strategy. We first perform the image enhancement for $\textbf{I}_{\rm DAB}$. We employ a threshold $t_{\rm intense}$ to obtain a intense-stained mask $\textbf{M}_{\rm intense}$. Then, another threshold $t_{\rm weak}$ ($<t_{\rm intense}$) is used to segment the image $\textbf{I}_{\rm DAB}$ to obtain a weakly/moderately brown-stained (DAB-stained) image mask $\textbf{M}_{\rm weak}$. Note that, this segmentation mask  $\textbf{M}_{\rm weak}$ also contains intense-stained regions.

\subsubsection{Skeletonization of Membrane}
We employ the skeletonize algorithm~\cite{zhang1984fast} to extract the contour map $\textbf{C}_{\rm weak}$ from the segmented mask $\textbf{M}_{\rm weak}$ to describe the cell membrane. We then analyze the skeleton by the contour connection. If the closed contour is the innermost contour, we mark the contour as a complete cell membrane, and incomplete otherwise. Too avoid perform the skeletonization for intense mask again, we obtain the skeleton (\ie, $\textbf{C}_{\rm intense}$) of $\textbf{M}_{\rm intense}$ by finding the intersection of $\textbf{C}_{\rm weak}$ and $\textbf{M}_{\rm intense}$. It is not strict, but the bias is not large.

\subsubsection{Mask Extraction}
We fill the enclosed skeleton in $\textbf{C}_{\rm weak}$  to generate an enclosed area mask, named $\textbf{M}_{\rm weak,enclosed}$ (Figure 5), and we define the set of coordinates of all foreground pixels in $\textbf{M}_{\rm weak, enclosed}$ as $\mathbb{P}_{\rm weak,enclosed}$. Similarly, we fill the enclosed area in $\textbf{C}_{\rm intense}$  to obtain $\textbf{M}_{\rm intense,enclosed}$, and we define the set of coordinates of all foreground pixels in $\textbf{M}_{\rm intense,enclosed}$ as $\mathbb{P}_{\rm intense,enclosed}$.

\begin{figure}[t]
	\centering
	\includegraphics[width=\textwidth]{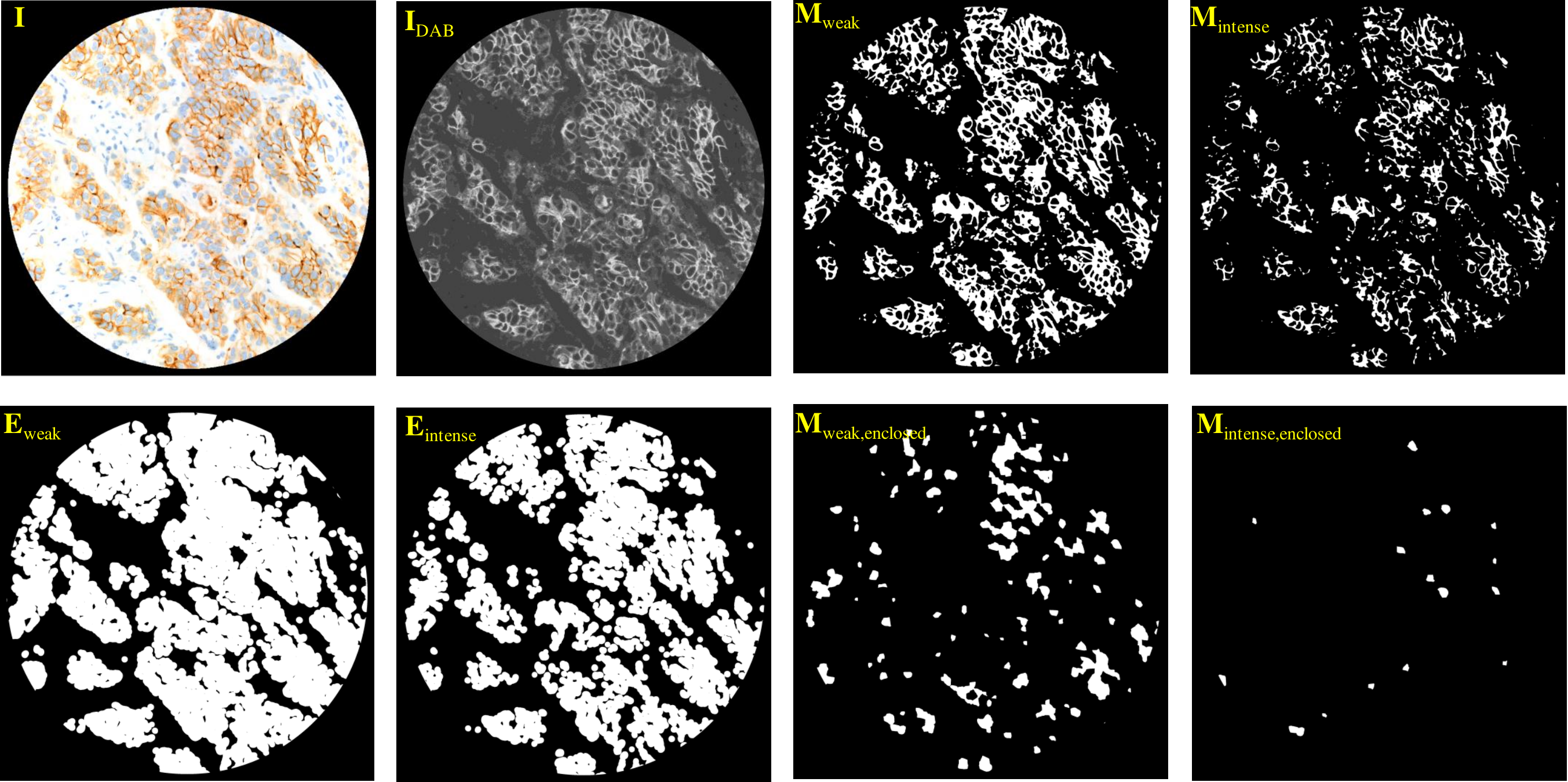}
	\caption{Extracted image masks for HER2 scoring. }
	\label{fig_masks}
\end{figure}

Besides extracting the two masks to represent enclosed area, we dilate the masks $\textbf{M}_{\rm weak}$ and $\textbf{M}_{\rm intense}$ with a kernel of $d$ to obtain the expanded masks of $\textbf{E}_{\rm weak}$ and $\textbf{E}_{\rm intense}$ respectively. The coordinates of all foreground pixels in $\textbf{E}_{\rm weak}$ and $\textbf{E}_{\rm intense}$ consist of two point sets $\mathbb{P}_{\rm weak}$ and $\mathbb{P}_{\rm intense}$ respectively. The kernel $d$ is a kind of distance parameter that is related to the radius of the cancer cells. The example masks are shown in Fig.~\ref{fig_masks}.

\subsection{Strategy-based Cell Classification}
To classify the detected tumor cells, it is necessary to quantify the cell staining (whether it is completely wrapped by intense, moderate/weak cell membranes). We recognize the cell types by a set of foreground coordinates (\ie, $\mathbb{P}_{\rm weak,enclosed}$, $\mathbb{P}_{\rm intense,enclosed}$, $\mathbb{P}_{\rm weak}$, $\mathbb{P}_{\rm intense}$) from multiple segmentation masks. We denote the set of all points in the entire image is $\rm U$.

\begin{enumerate}
  \item Intense \& complete tumor cells. The set of this type of cells is denoted as $\mathbb{T}_{\rm intense,complete}$, which is calculated as $\mathbb{D}_{\rm detected}\bigcap \rm \mathbb{P}_{intense,enclosed}$. The number of this category is marked as $card(\rm\mathbb{T}_{intense,complete})$.
  \item Intense \& incomplete tumor cells. The set of this type of cells is denoted as $\rm\mathbb{T}_{intense,incomplete}$, which is calculated as $\mathbb{D}_{\rm detected}\bigcap \rm C_{\mathbb{U}}\mathbb{P}_{intense,enclosed} \bigcap \mathbb{P}_{intense}$. The number of this category is marked as $card(\rm\mathbb{T}_{intense,incomplete})$.
  \item Moderate (weak) \& complete tumor cells. The set of this type of cells is denoted as $\rm\mathbb{T}_{weak,complete}$, which is calculated as $\rm \mathbb{D}_{detected}\bigcap C_{U}(\mathbb{T}_{intense,complete} \bigcup \mathbb{T}_{intense,incomplete}) \bigcap \mathbb{P}_{weak, enclosed}$. The number of this category is marked as $card(\rm \mathbb{T}_{weak,complete})$.
  \item Moderate (weak) \& incomplete tumor cells. The set of this type of cells is denoted as $\rm\mathbb{T}_{weak,incomplete}$, which is calculated as $\rm \mathbb{D}_{\rm detected}\bigcap C_{U}(\mathbb{T}_{intense,complete} \bigcup \mathbb{T}_{intense,incomplete} \bigcup \mathbb{T}_{weak,complete}) \bigcap \mathbb{P}_{weak, enclosed}$. The number of this category is marked as $card(\rm\mathbb{T}_{weak,incomplete})$.
  \item No staining tumor cells. The set of this type of cells is denoted as $\rm\mathbb{T}_{nostaining}$, which is calculated as $\rm\mathbb{D}_{\rm detected}\bigcap C_{U}(\mathbb{T}_{intense,complete} \bigcup \mathbb{T}_{intense,incomplete} \bigcup \mathbb{T}_{weak,complete} \bigcup \mathbb{T}_{weak,incomplete})$. The number of this category is marked as $card(\rm\mathbb{T}_{nostaining})$.
\end{enumerate}

\subsection{HER2 Scoring}
The HER2 scoring guidelines (such as breast cancer guidelines) are defined on the whole slide of HER2, and the microscope provides several FOVs selected by pathologists. Therefore, it is difficult to stitch these FOVs to a whole slide image. In view of this, we require the pathologists to select multiple typical FOVs that approximately represent the whole slide for HER2 scoring. Specifically, our HER2 scoring has the following steps:
\begin{enumerate}
  \item A radiologist is responsible for reading the whole slide and collecting multiple microscopic FOVs that typically include invasive cancer (because the HER2 score is defined on the staining of invasive cancer cells), such as 5-10 microscopic FOVs (both 20 and 40 power objectives can be selected).
  \item During the reading, our AI algorithm provides both quantitative counts and qualitative illustration of cell locations and membranes for each view in real-time.
  \item After selecting necessary typical FOVs, our system calculates the overall proportion of cells with different staining patterns for all FOVs.
  \item According to the scoring criteria in guidelines, we can determine the HER2 score. For example, if the proportion of intense\&complete cells exceeds 10\%, it will be a 3+ HER2 slide.
\end{enumerate}

\subsection{Implementation Details}
The size of image for each FOV is $3008\times3008$, with the pixel size of $0.212\times0.212 \mu m^2$ and $0.424\times0.424 \mu m^2$ for 40x and 20x respectively. The nucleus detection model is trained from $2927$ image patches (with the size of $2048\times 2048$) cropped from around 500 WSIs scanned in 40x magnification. We train the model for the pixel size of  $0.212\times0.212 \mu m^2$ and $0.424\times0.424 \mu m^2$, respectively. The data augmentation includes slightly scaling, rotation, flipping, grammar transform, and smoothing. We employ the LinkNet~\cite{chaurasia2017linknet} with Mean Square Error (MSE) loss for nucleus detection. Adam optimizer is selected with a learning rate of 0.001. In order to speed up the algorithm, we perform the nucleus detection in the original resolution, but the downsampled image ($1504\times1504$) for all the other processings.

\section{Results}
We quantitatively evaluate the nucleus detection and final HER2 scoring. The nucleus detection is evaluated in terms of Recall, Precision, and F1-score. The distance between detected nucleus and ground-truth nucleus less than $5\mu m$ is regarded as the true positive detections. For HER2 coring, we employ the evaluation for each view and whole slide, respectively. Note that, it is not preciseness to define the score for each view, but the scoring for each view can evaluate the algorithm exhaustively. Note that, it would be good to evaluate the classification performance of cells. However, it is challenging to manually annotate the cells with specific class labels.  Therefore, we directly evaluate the performance of HER2 scoring.

\subsection{Results for nucleus detection}
We evaluate the tumor nucleus detection performance on the independent dataset, which contains 20 images captured from 20x FOVs and 20 images captured from 40x FOVs. Note that, the FOVs do not include the DCIS regions. As shown in Table~\ref{table_nucleus}, the tumor nuclei could be detected in a very high recall for both 20x and 40x FOVs. Compared with the results in 20x FOVs, slightly better precision could be achieved for 40x FOVs, due to more detailed structural information could be captured. Overall, the tumor nucleus detection method using heatmap regression can achieve satisfactory detection performance and can distinguish the tumor cells from others, such as stromal cells and lymphocytes.

\begin{table}[!t]

\caption{Results for tumor nucleus detection}
\label{table_nucleus}
\centering
\begin{tabular}{c|c|c}
\hline
\hline
Method & 20x & 40x \\
\hline
Recall           & 0.92& 0.91 \\
Precision           & 0.73& 0.76 \\
F1-score           & 0.81& 0.82 \\
\hline
\hline
\end{tabular}
\end{table}
\subsection{Results for slide-level HER2 scoring}

We evaluate our system with 285 breast HER2 slides. The pathologists read each slide regularly and save necessary views (by pressing the foot pedal) for our AI system. To our knowledge, most of the existing HER2 scoring methods perform the classification task on the image patches (from WSI) or WSI. It is difficult to find a fair application scenario to compare different methods. As shown in Table~\ref{table_wholeslide}, we only report several methods for the classification of WSI in their studies. Compared with the existing methods, our microscope-based system achieves superior scoring performance on a much larger dataset.


\begin{table}[!t]

\caption{Classification results for whole slides.}
\label{table_wholeslide}
\centering
\begin{tabular}{c|c|c}
\hline
\hline
Method &Data&Accuracy \\
\hline
khameneh et al.~\cite{khameneh2019automated}             & 127 WSIs& 0.87 \\
Qaiser et al.~\cite{qaiser2019learning}            &86WSIs&0.79\\
Ours       & 285 slides & 0.95\\
\hline
\hline

\end{tabular}
\end{table}

%
%
%
\section{Discussion}

\subsection{Pathologist-AI Interaction}
Our system is integrated into the microscope that relies on the interactive FOV selection by pathologists. Taking advantage of the knowledge of pathologists, we can tackle the tough problem of distinguishing ductal carcinoma in situ (DCIS). On the one hand, the pathologists can avoid selecting FOVs having DCIS as well as inflammation regions. If it is difficult to avoid such confounding areas, we also provide a tool that can manually outline regions of interest. On the other hand, AI helps pathologists count cells accurately, which is almost impossible for pathologists to complete the counting task in multiple views. Because of the collaboration between pathologists and AI, the accurate and robust diagnoses could be achieved. Hopefully, precise quantization can help provide more explicit information for the following treatment.

\subsection{Thresholds for membrane segmentation}
Both the intense threshold $t_{intense}$ and moderate/weak threshold $t_{weak}$ were jointly adjusted to classify the 1000 FOVs, including challenging cases. Therefore, we select two thresholds that achieve the best classification performance. However, the staining style for IHC is not that consistent for using different staining reagents and types of equipment. In response to this, we have a tool bar for pathologists to further finetune these two thresholds subjectively to adjust various staining condition.

\subsection{Extension to have more specific quantization levels}
In this version of the HER2 scoring system, the moderate/weak staining membrane is regarded as a merged category for simplicity. It is easy to extend our method to classify the tumor cells into more specific categories to further differentiate the weak and moderate staining membrane. We have quantitatively evaluated the system with complex cell categories, there was no significant performance difference but slightly increasing the computational time.

\subsection{Extension to other cancer types}
Currently, we develop the algorithm according to the breast guidelines. Our method can be potentially extended to other cancer types, such as gastric and gastroesophageal junction cancer. However, the definition of HER2 scores may be different in different cancer guidelines. Slightly modification of the scoring criteria will be necessary.
\subsection{Computational cost}
We also analyze the computational cost of our HER2 scoring system. Since our system is integrated to the microscope, the efficiency is an important indicator for clinical application.  Table~\ref{table_cost} reports the computational cost for each module of our system. A CPU (Intel$^{\circledR}$ Xeon$^{\circledR}$ W-2133, 3.6GHz) and a GPU (Geforce RTX 2080, 8G) are employed in this experiment.
\begin{table}[!t]
\caption{Computational cost for each FOV in our HER2 scoring system.}
\label{table_cost}
\centering
\begin{tabular}{c|c|c|c}
\hline
\hline
Total &Nucleus Detection& Membrane Description & Cell Classification \\
\hline
0.6s   & 0.3s& 0.2s&0.1s \\
\hline
\hline

\end{tabular}
\end{table}

\section{Conclusion}
In this paper, we presented a HER2 scoring system based on the microscope, which can be integrated to pathologists' the routine workflow. Notably, we followed the HER2 scoring guidelines to perform the interpretation, making it easy to understand. Moreover, the pathologists can also help select informative FOVs and reduce the effect of confounding regions of DCIS. The validation on a set of 285 HER2 slides showed the effectiveness of our system. Hopefully, our method can improve diagnostic accuracy and both the inter-/intra-observer reliability.

\bibliographystyle{unsrt}
\bibliography{sample}

\end{document}